\documentclass[conference]{IEEEtran}

\usepackage{cite}

\usepackage{multicol}
\usepackage{graphicx}
\usepackage{multirow}
\usepackage{framed}
\usepackage{verbatim}
\usepackage{subfigure}
\usepackage{url}
\usepackage{textcomp}
\usepackage{todonotes}
\usepackage{amsfonts}
\usepackage{amssymb}
\usepackage{booktabs}
\usepackage{color}
\usepackage{array}
\usepackage{amstext}
\usepackage{flushend}

\usepackage[ruled, linesnumbered]{algorithm2e}

\newcommand{\aspas}[1]{{``#1''}}

\newcommand{\tcode}[1]{\texttt{\footnotesize{#1}}}

\newcommand{\repo}[1]{{\footnotesize{\sf{#1}}}}

\definecolor{gray}{RGB}{90,90,90}
\newcommand\sbar[2]{{\color{gray}\rule{\dimexpr 1cm * #1 / #2}{5pt}}}

\begin{document}

\title{A Novel Approach for Estimating Truck Factors}

\author{\IEEEauthorblockN{Guilherme Avelino\IEEEauthorrefmark{1}\IEEEauthorrefmark{2}, 
 Leonardo Passos\IEEEauthorrefmark{3},
 Andre Hora\IEEEauthorrefmark{1} and
 Marco Tulio Valente\IEEEauthorrefmark{1}
 }
\IEEEauthorblockA{\IEEEauthorrefmark{1}ASERG Group, Department of Computer Science (DCC)\\
 Federal University of Minas Gerais (UFMG), Brazil\\
 Email: \{gaa, mtov, hora\}@dcc.ufmg.br}
\IEEEauthorblockA{\IEEEauthorrefmark{2} 
    Department of Computing (DC)\\  
    Federal University of Piaui (UFPI), Brazil}
\IEEEauthorblockA{\IEEEauthorrefmark{3}University of Waterloo, Canada\\
 Email: lpassos@gsd.uwaterloo.ca}
}
\maketitle

% . . . . . . . . . . . . . . . . . . . . . . . . . . .

\begin{abstract}
Truck Factor (TF) is a metric proposed by the agile community as a tool to 
identify concentration of knowledge in software development environments. It 
states the minimal number of developers that have to be hit by a truck (or 
quit) before a project is incapacitated. In other words, TF helps to measure 
how prepared is a project to deal with developer turnover. Despite its clear 
relevance, few studies explore this metric. Altogether there is no 
consensus about how to calculate it, and no supporting evidence backing
estimates for systems in the wild. To mitigate both issues, we propose 
a novel (and automated) approach for estimating TF-values, which we execute 
against a corpus of 133 popular project in GitHub. We later survey
developers as a means to assess the reliability of our results. 
Among others, we find that the majority of our target systems (65\%) have TF 
$\leq$ 2. Surveying developers from  67 target systems provides  
confidence towards our estimates; in 84\% of the valid answers we collect, 
developers agree or partially agree that the TF's  authors are the main authors 
of their systems; in 53\% we receive a positive or partially positive answer 
regarding our estimated truck factors.
\end{abstract}

% . . . . . . . . . . . . . . . . . . . . . . . . . . .


% . . . . . . . . . . . . . . . . . . . . . . . . . . .

\section{Introduction}

A system's truck factor (TF) is defined as \aspas{\textit{the number of people
on your team that have to be hit by a truck (or quit) before the project is in serious 
trouble}}~\cite{Williams2003}. Systems with a low truck factor spot strong dependencies towards 
specific personnel, forming knowledge silos among developer teams. If such 
knowledgeable personnel abandon the project, the system's lifecycle is seriously 
compromised, leading to delays in launching new releases, and ultimately to 
the discontinuation of the project as whole. To prevent such issues, 
comprehending a system's truck factor is a crucial mechanism.

Currently, the existing literature defines truck factor loosely. For the most 
part, there is no formal definition of the concept, nor means to estimate it. 
The only exception we are aware of stems from the work of Zazworka et 
al.~\cite{Zazworka2010}. Their definition, however, as well as follow-up 
works~\cite{Ricca2010,Ricca2011}, is not backed by empirical evidence from 
real-world software systems. Stated otherwise, TF-estimates, as calculated by 
Zazworka's approach, lack reliability evidence from systems in the wild.

Our work aims to improve the current state of affairs by proposing a novel 
approach for estimating truck factors, backed up by empirical evidence to 
support the estimates we produce. In particular, we define an automated 
workflow for TF-estimation for which we apply to a target corpus comprising 133 systems in GitHub. In total, such systems have over 373K 
files and 41 MLOC; their combined evolution history sums to over 2 million 
commits. By surveying and analyzing answers from 67 target systems, we evidence 
that in 84\% of valid answers developers agree or partially agree that the TF's 
authors are the main authors of their systems; in 53\% we receive a positive or 
partially positive answer regarding our estimated truck factors. 

From our work, we claim the following contributions:

\begin{enumerate}

%\item  Using authorship measurements, we reveal the ratio of authors in a 
%system, the inequality  of authorship distributions, the frequency of single 
%authorship files, and the  impact of a system's size and programming language  
%in the authorship results (Section~\ref{sec:authorship-results}).\\[-0.4cm]

\item A novel approach for estimating a system's truck factor, as well as a 
publicly available supporting 
tool.\footnote{https://github.com/aserg-ufmg/Truck-Factor}

\item An estimate of the truck factors of 133 GitHub systems. All our data 
is publicly available for external 
validation,\footnote{\url{http://aserg.labsoft.dcc.ufmg.br/truckfactor}} 
comprising the largest dataset of its kind.

\item Empirical evidence of the reliability of our truck factor estimates, as a 
product of surveying the main contributors of our target systems.
From the survey, we report the practices that developers argue as 
most useful to overcome a truck factor event.
\end{enumerate}

We organize the remainder of the paper as follows. In 
Section~\ref{sec:python} we present a concrete example of 
truck factor concerns in the early days of Python development. In 
Section~\ref{sec:approach} we present our novel approach for truck factor 
estimation, detailing all its constituent steps. Next, Section 
\ref{sec:methodology} discusses our validation methodology, followed by the 
truck factors of our target systems in Section \ref{sec:estimates}. We proceed 
to present our validation results from a survey with developers 
(Section~\ref{sec:validation:tf}), further discussing results in 
Section~\ref{sec:discussions}. We argue about possible threats in 
Section~\ref{sec:threats}. We present the related work in 
Section~\ref{sec:relatedwork}, concluding the paper in 
Section~\ref{sec:conclusion}.

% . . . . . . . . . . . . . . . . . . . . . . . . . . .

\section{Truck Factor: An Example from the Early Days of Python}
\label{sec:python}

\noindent{\emph{``What if you saw this posted tomorrow: Guido's unexpected 
death has come as a shock to us all. Disgruntled members of the Tcl mob are 
suspected, but no smoking gun has been found..."}}---Python's mailing list 
discussion,  
1994.\footnote{http://legacy.python.org/search/hypermail/python-1994q2/1040.html
}\vspace{0.2cm}

Years before the first discussions about truck factor in eXtreme programming 
realms,\footnote{\url{
http://www.agileadvice.com/2005/05/15/agilemanagement/truck-factor/}} this 
post illustrates the serious threats of knowledge concentration in software 
development. By posting the fictitious news in Python's mailing list, the 
author, an employee at the National Institute of Standards and Technology/USA, 
wanted to foster the discussion of Python's fragility resulting from its strong 
dependence to its creator, Guido van Rossum:

\noindent{\emph{``I just returned from a meeting in which the major objection to 
using Python was its dependence on Guido. They wanted to know if Python
would survive if Guido disappeared. This is an important issue for
businesses that may be considering the use of Python in a product."}}

Fortunately, Guido is alive. Moreover, Python no longer has a truck factor 
of one. It has grown to be a large community of developers and the 
fifth most popular programming language in 
use.\footnote{\url{http://www.tiobe.com/index.php/tiobe_index}} 
However, the message illustrates that Python was, at least by some, considered
a risky project. As knowledge was not collective among its team members, 
but rather concentrated in a single ``hero", in the absence of the 
latter, discontinuation was a real threat, or at minimum, something that 
could cause extreme delays. Projects with low truck factor, as Python was in 
1994, face high risk adoption, discouraging their use. 

To facilitate decision making, it is crucial to metrify the risks of project 
failure due to personnel lost. This information serves not only business 
managers assessing early technology adoption risks, but also maintainers and 
project managers aiming to identify early knowledge silos among  
development teams. Timely action plans can then be devised as a 
means to prevent long-term failure. In that direction, reliable estimations of a 
project's truck factor is a must.

% . . . . . . . . . . . . . . . . . . . . . . . . . . .

\section{Proposed Approach}
\label{sec:approach}

We calculate the truck factor of a target system by processing its evolution 
history. We assume the latter to be managed by a version control 
system, in addition to having access to a local copy of the repository of the 
target subject. 

\begin{figure}[!t]
    \centering
    \includegraphics[width=\linewidth]{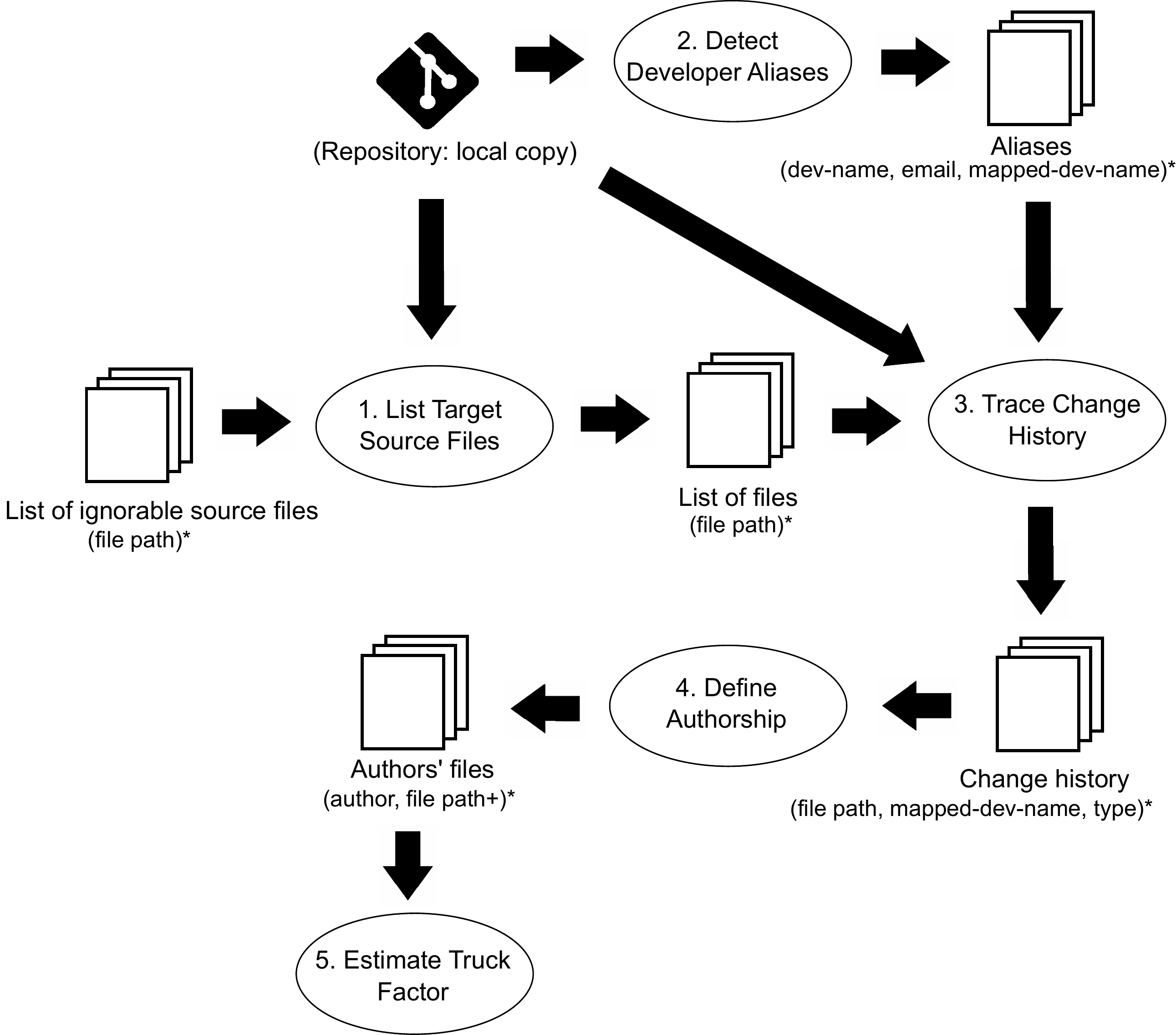}
    \caption{Proposed approach for truck factor calculation}
    \label{fig:process}
\end{figure}

Our approach comprises five major steps---see Figure~\ref{fig:process}. 
Step 1 checkouts the latest point in the commit history, listing all the source 
files therein. Step 2 handles possible aliases among developers, \emph{i.e.}, 
cases where a single developer has multiple Git users. Step 3 traces the 
history of each source code file. From such traces, step 4 defines the authors 
in the system, as well as their authored files. Authorship, in this case, is 
not a strict notion. Stated otherwise, authorship is not a matter of
who creates a file. Rather, authorship is a statement of who will be able to 
maintain a file from the latest system snapshot onward. This may comprise the 
creator of the file (original author), as well as other developers 
(co-authors) who significantly contributed with changes to a 
file after its creation. With the list of authors and their authored files, step 
5 estimates the truck factor of the entire system.

We realize the given process to automatically estimate the truck factor of 
projects whose evolution is managed by Git. In the following, we detail the 
realization of each step.

\subsection{Realization}

\noindent\textit{Step 1: List Target Source Files.} To obtain the list of 
target files, we first switch to the master branch of the target repository, 
checking out its latest commit. We then enumerate the path of all source files 
of the given snapshot, excluding all other file types (\emph{e.g.}, files representing 
documentation, images, examples, etc), as well as the files listed in the 
ignorable source file list, given as input. We also discard source  
files associated with third-party libraries  (\emph{i.e.}, files that are not developed 
in the system under analysis). Our decision is conservative. An existing survey 
from JavaOne'14\footnote{\url{http://tinyurl.com/javaone14-survey}} reports 
that nearly two-thirds of polled senior IT professionals have
Java applications with half of their code coming from third-party sources. 
Thus, if developers store third-party code in the system's main Git repository 
(\emph{e.g.}, as backup, to facilitate build, etc), and third-party code is as large 
as the poll suggests, truck factor estimates are likely to be significantly 
affected. 

To exclude third-party code, one must be able to identify it in the first 
place. As such, we employ 
Linguist,\footnote{\url{https://github.com/github/linguist}} an opensource 
tool from GitHub. Linguist is actively developed, and it is constantly being 
updated by the GitHub community to include new pattern matching rules to 
identify third-party file names. Linguist's original goal is to
detect the programming language of GitHub projects---as in our 
case, this is sensitive to external code.\\[0.2cm]
%
% Linguist, however, is architecture and system agnostic. Thus, it does not 
% account for the presence of user-pluggable components, which are recurrent in 
% plugin-based software systems (e.g., Eclipse). As with third-party code, 
% plugins also influence a system's truck factor. To exclude such files
% when later estimating a system's truck factor, step 1 requires a list of all 
% files implementing plugins. This is the only part of our process whose input has 
% to be manually defined. We argue, however, that such a listing can be easily 
% produced, provided plugins are kept within specific folders of a system's 
% source code.\footnote{If a Java-based system stores its plugins inside
% a \tcode{plugins} folder or any of its subfolders, all Java source files 
% can be listed by simply issuing 
% \tcode{find plugins -type f -name *.java}} \\[0.2cm]
%
\newcommand{\bl}{\textasciicircum}
\newcommand{\bs}{\textbackslash}
\newcommand{\tl}{\textless}
\newcommand{\tg}{\textgreater}
\newcommand{\el}{\textdollar}
\newcommand{\sm}{\textquotesingle}
\noindent\textit{Step 2: Detect Developer Aliases.} Each user in Git is a pair 
(\textit{dev-name}, \textit{email})---\emph{e.g.},~(\aspas{Bob Rob}, 
\aspas{bob.rob@example.com}). It happens, however, that a single 
developer may be associated with many Git user accounts, leading to \textit{developer 
aliases}. As such, this step first extracts all dev-names and 
their associated emails.\footnote{For instance, by issuing 
\tt{\footnotesize{git 
log | grep "Author:" | sed 
\sm{s}/\bl.*:\bs{s}\bs{+}//;s/\bs{s}\bs{+}\tl/;/;s/\tg\el//\sm | sort | uniq}}} 
From the listing, we group users with the same email, but possibly with 
different associated dev-names. Additionally, we unify similar dev-names, but 
with different emails. In the latter case, we employ the Levenshtein 
distance~\cite{Navarro2001}, with a threshold of at most one. This means that 
we allow at most one single-character insertion, deletion or substitution to 
claim two different dev-names as referring to the same developer. For example, 
if \aspas{Bob.Rob} and \aspas{Bob Rob} are two dev-names, we consider them as 
the same developer, as a single substitution is needed to make their names 
identical. As a result of alias detection, step 2 outputs a mapping from Git 
users to a single developer name (\textit{mapped-dev-name}).\\[0.2cm]
\noindent\textit{Step 3: Trace Change History.} This step traces the evolution 
history of each target file, taking as input the results of the previous two 
steps. 
%To perform the tracing, we employ the \textit{follow} facility of Git: 
%if \tcode{target.src} is a file to trace, \tcode{git log --follow -p
%target.src} retrieves all commits and associated patches modifying 
%\tcode{target.src},  including possible renames, as well as the original file 
%addition. 
To perform the tracing, we collect the system's commits using the \tcode{git 
log --find-renames} command. This command returns all commits of a repository 
and identifies possible file renames.
We process each commit, extracting three pieces of information: (i) the path 
of the file we are collecting the trace; (ii) the \textit{mapped-dev-name} of 
the developer performing the change; and (iii) the type of the change---file 
addition, file modification, or file rename.\\[0.2cm]
\noindent\textit{Step 4: Define Authorship.} Given the change traces of each 
file in the target snapshot of the project at hand, this step defines the 
author list of each file. Different alternatives could be used 
as a means for determining 
authorship---\emph{e.g.},~\cite{Anvik2006,Minto2007,Schuler2008,Hattori2009,Fritz2010,
Fritz2014}.
Among those, we chose the
\textit{degree-of-authorship} (DOA) metric~\cite{Fritz2010,Fritz2014}, which we 
normalize after calculation. Given a file $f$ 
with path $f_p$, the degree-of-authorship of a developer $d$ whose Git user has 
been mapped to  $m_d$ is given by:

\[
\label{eq:DOA}
\begin{array}{l}
\mathit{DOA}(m_d, f_p) =  3.293 + 1.098 \times \mathit{FA}(m_d, f_p) + 
  0.164 \times \\ 
  ~~~~~~~~~~~~~~~
  \mathit{DL}(m_d, f_p) - 0.321 \times \ln(1 + \mathit{AC}(m_d, f_p))
\end{array}
\]

\noindent From the equation, DOA depends on three 
factors: (i) first authorship (FA): if $m_d$ originally created $f$, 
$\mathit{FA}$ is 1; otherwise it is 0; (ii) number of deliveries (DL): number 
of changes in $f$ made by $m_d$; and (iii) number of acceptances (AC): number 
of changes in $f$ made by any developer, except $m_d$. 

The model assumes FA as
the strongest predictor of authorship. Recency information (DL) 
positively contributes to authorship, but with less importance. 
In contrast, other developers' changes (AC) decrease one's DOA, although at a 
slower pace. The weights we use in the DOA model stem from empirical 
experiments performed elsewhere~\cite{Fritz2014}. 

DOA has three major advantages in comparison to other approaches: (i) assuming 
the weights of the model to be general, DOA does not require a training 
set; 
%(ii) DOA considers all the evolution history of a file, from its creation 
% %to all further modifications. By accounting all the changes made by 
%different %developers in a given file, results are likely to be more accurate 
%than approaches consider only  single-snapshot based approaches; 
(ii) DOA does not require monitoring editing activities as 
developers maintain different files (\emph{e.g.}, as in~\cite{Hattori2009}); (iii) 
instead of considering all developers changing a file as its authors (\emph{e.g.}, as 
in~\cite{Zazworka2010}), DOA weights contributions differently, 
accounting for both changes of a developer in a file (increases DOA), as 
well as the changes performed by others (decreases DOA).

Once we know the DOA-values of all files changed by previously mapped 
developers, we proceed to normalize results. A normalized DOA-value ranges 
from 0 to 1. We grant 1 to the developer with the highest absolute DOA among 
all developers that worked on $f$; altogether, we consider a developer as an 
author of a file if its resulting normalized DOA is greater than a threshold 
$k$ and its absolute DOA is not lower than a value $m$. 
Currently, $k$ and $m$ stands as configurable parameters in our approach. As we 
show in Section~\ref{sec:setup}, $k = 0.75$ and $m = 3.293$ seem to provide good results. As a result of  this step, we output a list of associations from 
authors (\textit{mapped-dev-names}) to their related authored files.\\[0.2cm]
\noindent\textit{Step 5: Estimate Truck Factor.} Taking a list $A$ of authors 
(\textit{mapped-devs}) and their associated authored files (one or more file 
paths), this step estimates the system's truck factor. Our estimation relies on 
a coverage assumption: a system will face serious delays or will be likely 
discontinued if its current set of authors covers less than 50\% of the current 
set of files in the system. Following such assumption, our truck factor 
estimation algorithm implements a greedy heuristic---see 
Algorithm~\ref{alg:tf}. 
Starting with a truck factor of zero, we iterate over the authors' file list $A$
(lines~\ref{alg:while:start}--\ref{alg:while:end}), verifying at each iteration 
whether the current authors' coverage is below 0.5 (line~\ref{alg:coverga}). If 
so, we stop the iteration---maintenance is likely to be hampered; otherwise, we 
remove the top author from $A$ (line ~\ref{alg:rm}), increasing truck factor by 
one (line \ref{alg:tf:inc}). The top author in a given iteration is the 
\textit{mapped-dev} authoring the highest number of files in $A$.\footnote{This 
is obtained by finding the entry 
$e_i = (a_i, \textit{filepath-list}_i) \in A$ s.t. 
$\nexists~e_j = (a_j, \textit{filepath-list}_j) \in A \land e_j \neq e_i 
\land 
|\textit{filepath-list}_j| > |\textit{filepath-list}_i|$. If there exist more 
than one top author, we just take the first one we find.} 
Whenever $A$ shrinks, another iteration follows, provided $A$ 
is not empty. This process continues until $A$ becomes empty or coverage is 
less than $0.5$.

\SetAlFnt{\footnotesize\sffamily}
\begin{algorithm}[t]  
    \SetAlgoLined
    \KwIn{List of authors' files $A$} 
    \KwOut{System truck factor}
    \Begin{
        $F$ $\gets$ \textit{getSystemFiles}($A$)\; \label{alg:files}
        \textit{tf} $\gets$ 0\;
        \While{$A$ $\neq$ $\emptyset$}{ \label{alg:while:start}            
          \textit{coverage} $\gets$ \textit{getCoverage}($F$, 
$A$)\;
            \If{coverage $<$ 0.5}{\label{alg:coverga}
                \textbf{break}\;    
            }
            $A$ $\gets$ \textit{removeTopAuthor}($A$)\;\label{alg:rm}
            \textit{tf} $\gets$ \textit{tf} + 1; \label{alg:tf:inc}           
        } \label{alg:while:end}
        \textbf{return} \textit{tf}\;	
    }%endBegin
    \caption{\textsc{Truck factor algorithm.}}
    \label{alg:tf}
\end{algorithm}

% . . . . . . . . . . . . . . . . . . . . . . . . . . .

\section{Validation Methodology}
\label{sec:methodology}

To validate our approach, we select 133 systems from GitHub. For 
each target system, we estimate its truck factor. 

This section details our corpus selection and how we setup our approach for 
estimating truck factors for our chosen subjects. We also discuss how we survey 
developers as a means to validate our estimates and get further insights.

% , as well as the experimental 
% setups for assessing the performance of the proposed approach when estimating 
% the truck factors of all 133 target projects.

\subsection{Selection of Target Subjects} 

To select a target set of subjects, we follow a procedure similar to other 
studies investigating 
GitHub~\cite{Yamashita2015,Gousios2014,Kalliamvakou2014b,Ray2014}.
First, we query the programming languages with the largest number of 
repositories in GitHub. We find six main languages ($L$): 
JavaScript, Python, Ruby, C/C++, Java, and PHP. We then select the 100-top most 
popular repositories within each target language. Popularity, in this 
case, is given by the number of times a repository has been starred by GitHub 
users. Considering only the most popular projects in a given language 
($S_\ell$), we remove the systems in the first quartile ($Q_1$) of the 
distribution of three metrics, namely number of developers ($n_d$), number of 
commits ($n_c$), and number of files ($n_f$). After filtering out subjects in 
$Q_1$,  we compute the intersection of the remaining sets. From the previous 
steps, we get an initial set of prospective subjects $T^0$. Formally, 
\[
  \displaystyle
  T^0 = \bigcup_{\ell \in L} 
     T^0_{n_d}(\ell) \cap 
     T^0_{n_c}(\ell) \cap 
     T^0_{n_f}(\ell)
\]
\noindent where 
\[
 \begin{array}{l}
  T^0_{n_d}(\ell) = S_\ell - Q_1(n_d(S_\ell))),~T^0_{n_c}(\ell) = S_\ell - 
Q_1(n_c(S_\ell))),  \\
  T^0_{n_f}(\ell) = S_\ell - Q_1(n_f(S_\ell)))
  \end{array}
\]

From $T^0$, we determine a new subset $T^1$ including only the systems whose 
repositories stem from a correct migration to GitHub. Specifically, we remove 
systems with more than 50\% of their files added in less than 20 commits---less 
than 10\% of the minimal number of commits we initially considered. This 
evidences that a large portion of a system was developed using another 
version control platform and the migration to GitHub could not preserve the 
original version history. From the resulting set of prospective 
subjects ($|T^1| = 135$), we manually inspect the documentation in each 
repository to identify and eliminate duplicate subjects. Our inspection shows
\repo{raspberrypi/linux} and \repo{django/django-old} as duplicate 
cases. The first, despite not being a fork, is very similar to 
\repo{torvalds/linux}; in fact, it is a clone of the Linux kernel, with 
extensions 
supporting RaspberryPi-based boards. The second is an old version of a 
repository already in $T^1$.

\begin{figure}[tb]
  \center      
  \subfigure[Developers]{\includegraphics[width=0.22\linewidth]
      {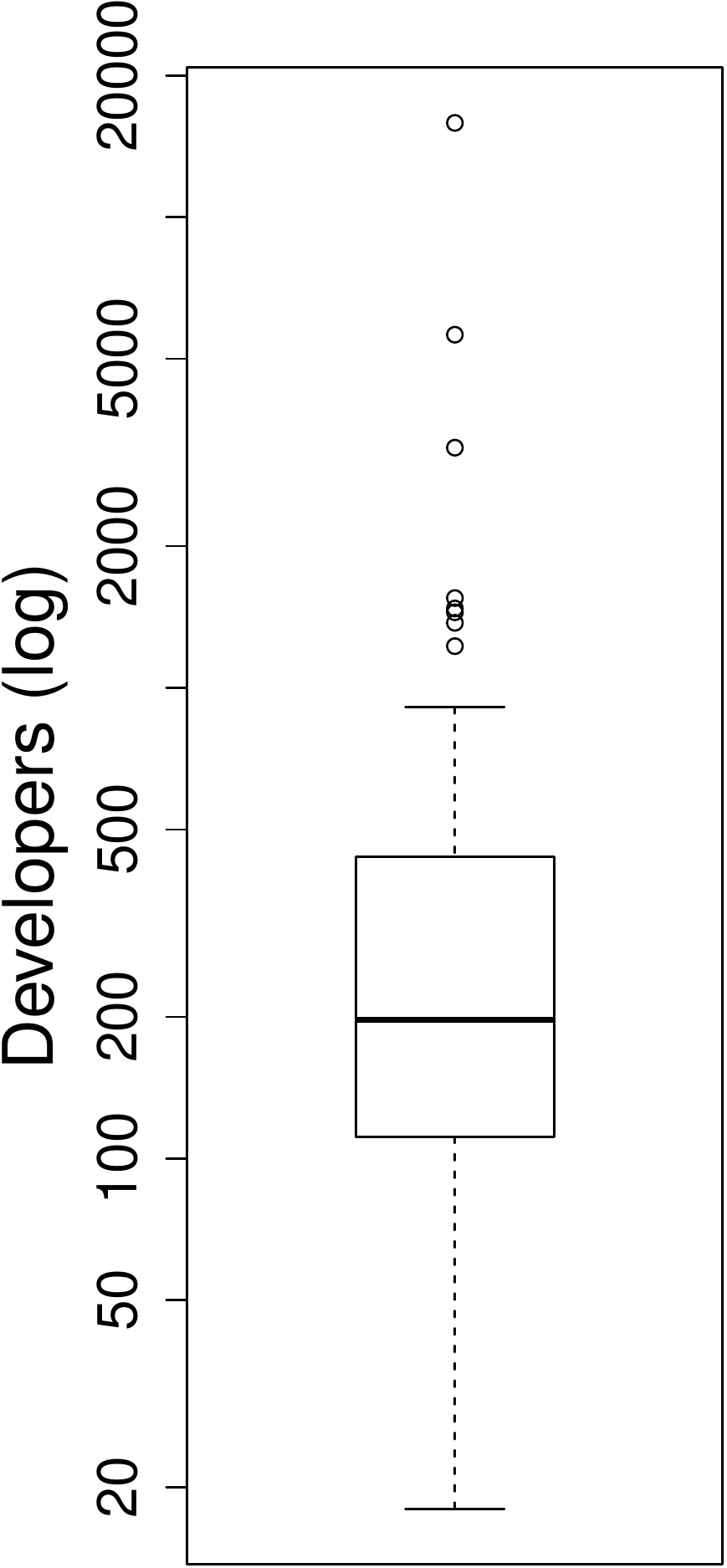}
  \label{fig:box-dataset:devs}}
\subfigure[Commits]{\includegraphics[width=0.22\linewidth]
    {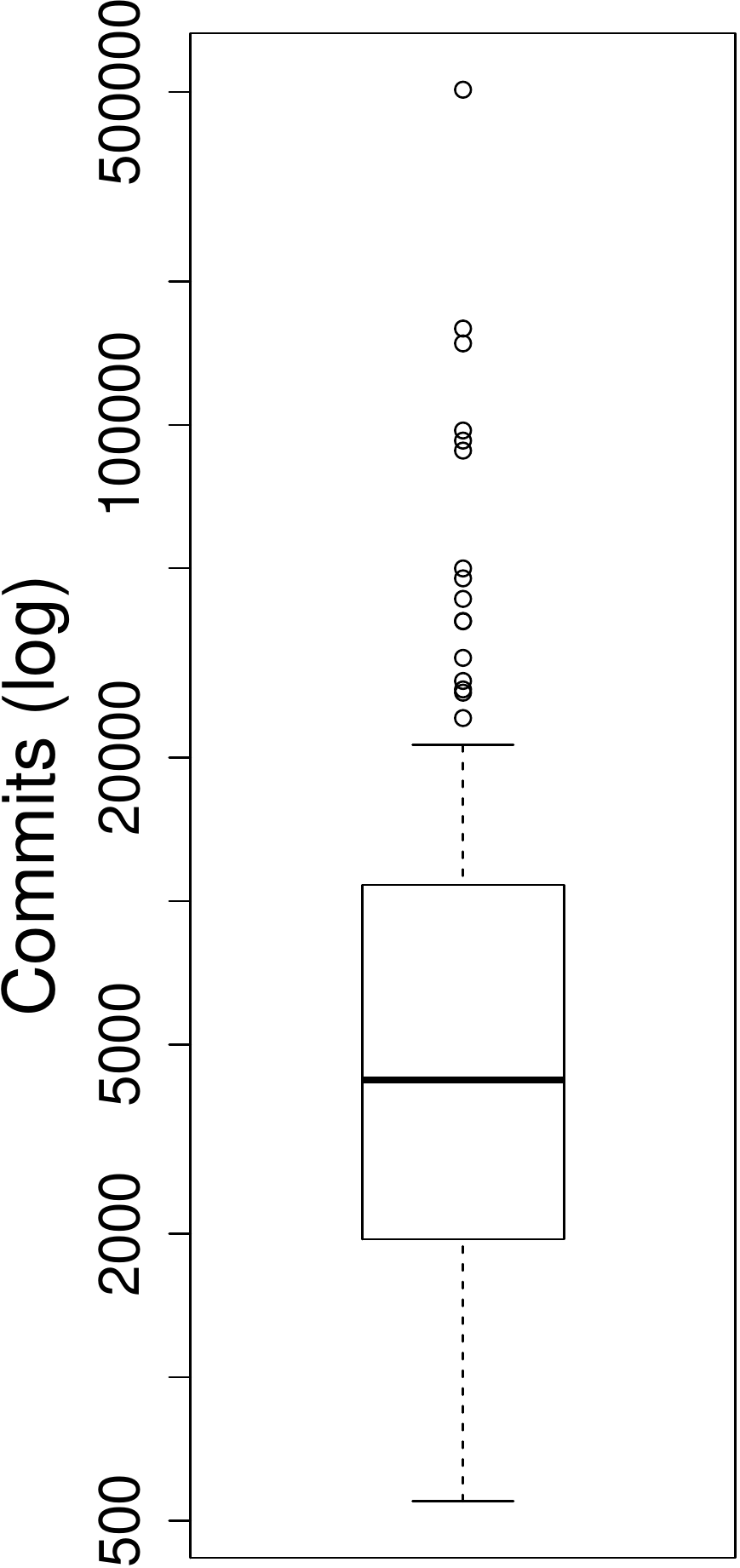}
    \label{fig:box-dataset:commits}}
\subfigure[Files]{\includegraphics[width=0.22\linewidth]
    {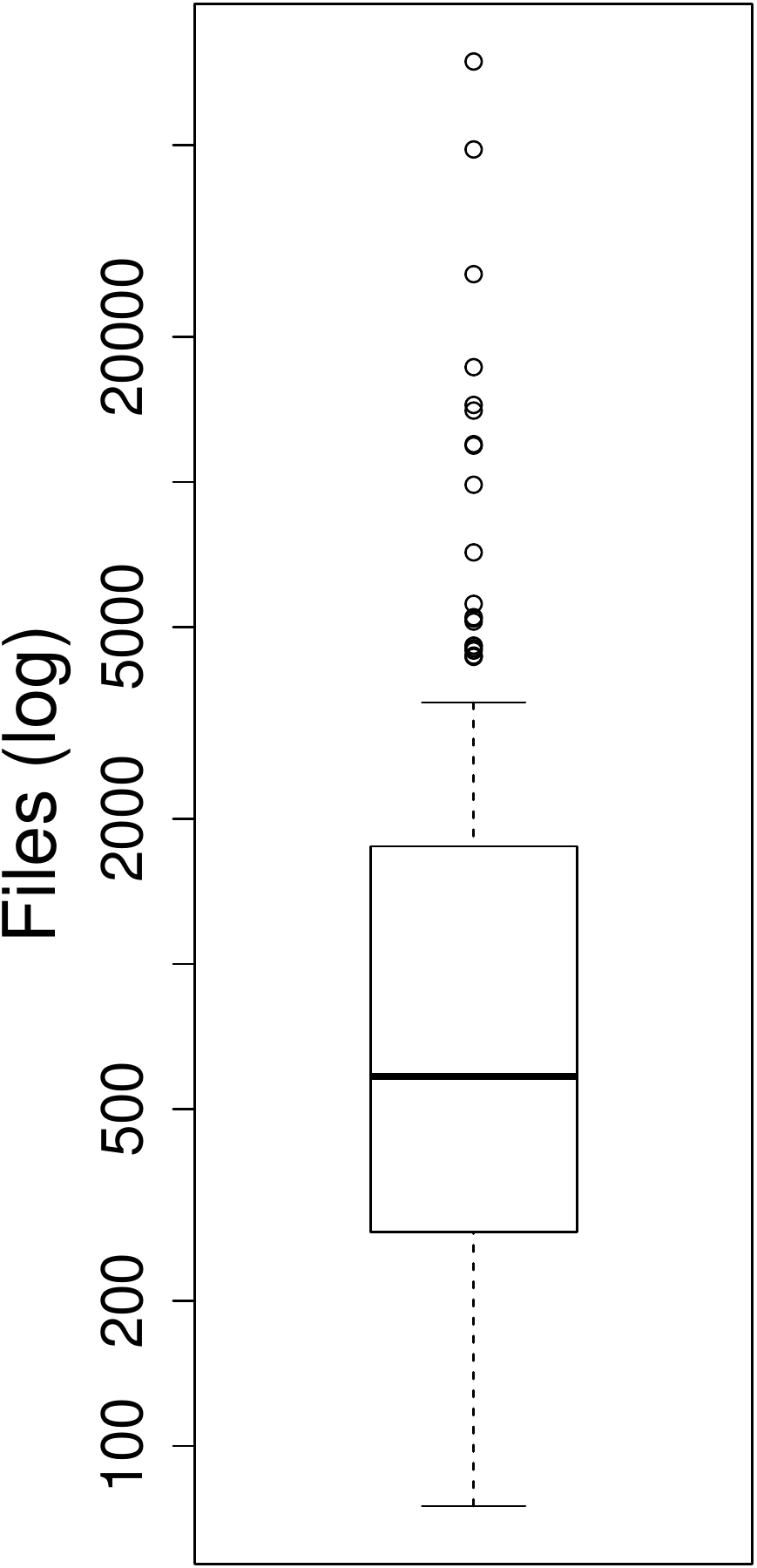}
    \label{fig:box-dataset:files}}
\subfigure[LOC]{\includegraphics[width=0.22\linewidth]
    {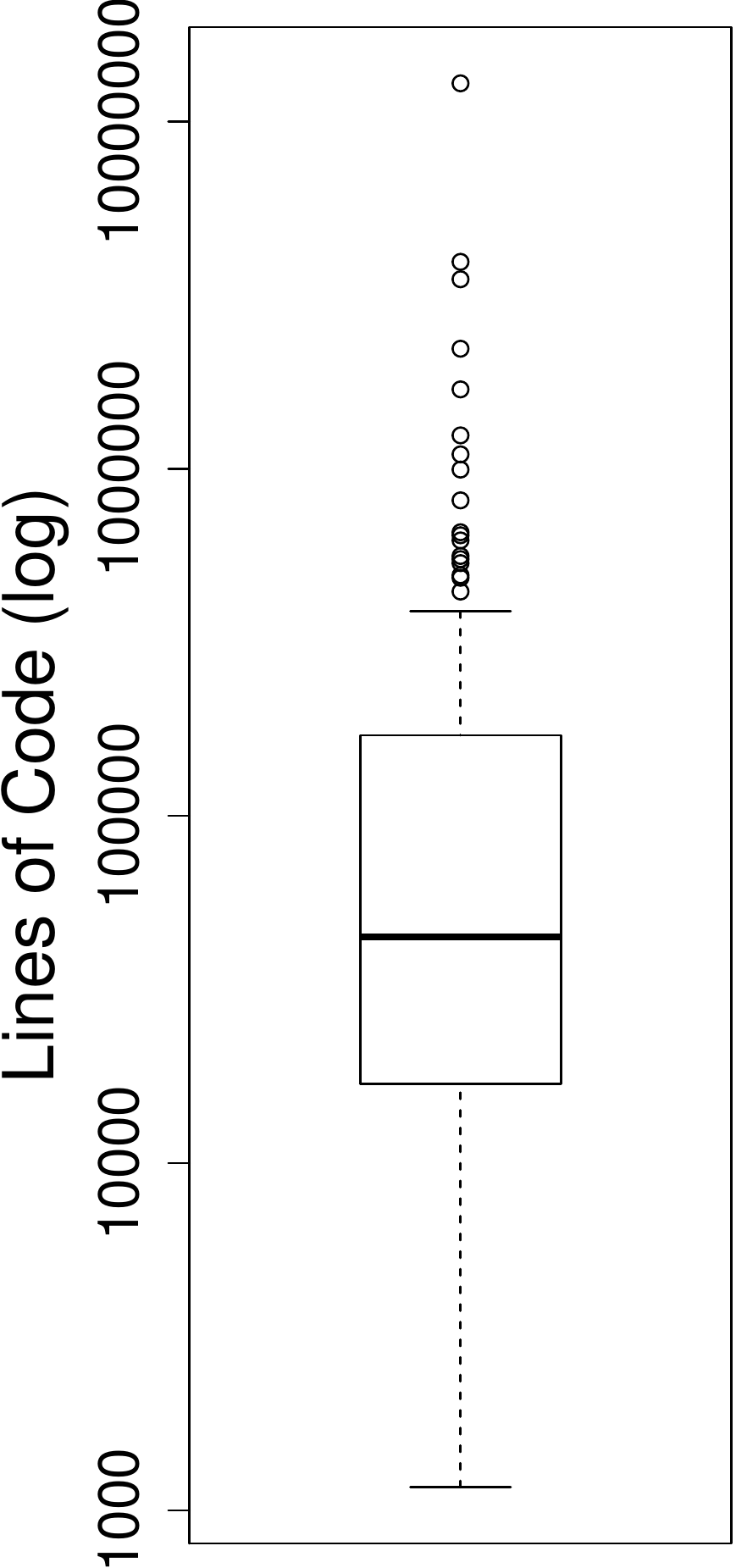}
    \label{fig:box-locs:devs}}
  \caption{Target subjects}
  \label{fig:box-dataset}
\end{figure}

After excluding \repo{raspberrypi/linux} and \repo{django/django-old}, we are 
left with 133 subjects ($T^2$), which represent the most important systems per 
language in GitHub, implemented by teams with a considerable number of active 
developers and with a considerable number of files. Table~\ref{tab:datasetinfo} 
summarizes the characteristics of the repositories of our chosen subjects. Ruby 
is the language with more systems, 33 in total. The programming language with 
less systems is PHP, with 17 projects. Accounting all our chosen 
subjects, their latest snapshots accumulate over 373K files and 41 MLOC; their 
combined evolution history sums to over 2 million commits. Our targets also 
have a large community of contributors, accumulating to over 60K developers.
\begin{table}[!t]
  \centering
  \caption{Target repositories}
  \label{tab:datasetinfo-reduced}
  \resizebox{\linewidth}{!}{
  \begin{tabular}{lrrrrr}
  \toprule
  \multicolumn{1}{c}{\bf Language} & 
  \multicolumn{1}{c}{\bf Repos} & \multicolumn{1}{c}{\bf Devs} & 
  \multicolumn{1}{c}{\bf Commits} & \multicolumn{1}{c}{\bf Files}  & 
  \multicolumn{1}{c}{\bf LOC}  \\
  \midrule      
  JavaScript & 22  & 5,740  & 108,080 & 24,688  & 3,661,722  \\
  Python     & 22  & 8,627  & 276,174 & 35,315  & 2,237,930  \\
  Ruby       & 33  & 19,960 & 307,603 & 33,556  & 2,612,503  \\
  C/C++      & 18  & 21,039 & 847,867 & 107,464 & 19,915,316 \\
  Java       & 21  & 4,499  & 418,003 & 140,871 & 10,672,918 \\
  PHP        & 17  & 3,329  & 125,626 & 31,221  &2,215,972   \\
  \textbf{Total}   & \textbf{133}  & \textbf{63,194}     
                   & \textbf{2,083,353} & \textbf{373,115} 
                   & \textbf{41,316,361} \\      
   \bottomrule
   \end{tabular}
  }
 \label{tab:datasetinfo}
\end{table}
Figures \ref{fig:box-dataset:files}--\ref{fig:box-locs:devs} depict each 
distribution.

% For files 
% (Figure~\ref{fig:box-dataset:files}), the first, second, and third quartiles 
% are 269, 574, and 1,750, respectively. The top-3 systems with the highest 
% amount of fil \repo{JetBrains/intellij-community} (74K files), 
% \repo{torval\-ds/\-li\-nux} (49K files), and 
% \repo{android/plat\-form\_fra\-me\-wor\-ks\-\_base} (27K 
% files). \todo[inline]{For LOC, ...} 
% %
% As for commits (Figure~\ref{fig:box-dataset:commits}), quartiles are
% 1,950, 4,210, and 10,800. The top-3 systems with more commits are 
% \repo{torvalds/linux} (505.6K commits), 
% \repo{android/plat\-form\_fra\-me\-works\_base} 
% (159.3K commits), and \repo{Jet\-Brains/in\-tel\-lij-community} (148.5K 
% commits).
% %
% The first, second and third quartiles of the number of developers 111, 197, and 
% 437, respectively. The top-3 systems with more developers are 
% \repo{torvalds/linux} (15.8K developers), \repo{Homebrew/homebrew} (5.6K 
% developers), and \repo{rails/rai\-ls} (3.2K developers).
%
%
\subsection{Setting up Inputs}
\label{sec:setup}

Our approach requires as input a listing of ignorable source files of 
a system, in addition to a tweak of the DOA thresholds ($k$ and $m$). 
Next, we detail how we set both inputs.\\[0.2cm]
\noindent\textit{List of Ignorable Source Files.} To create the list of 
ignorable files, we manually inspect the first two top-level directories in 
each target repository, seeking to find third-party libraries undetected by 
Linguist. Also, as Linguist is architecture and system agnostic, we 
look for plugin-related code in systems with a plugin-based architecture.
As with third-party code, plugins may highly influence a system's truck factor. 
For instance, in the Linux kernel, driver plugins are the most common feature 
type~\cite{Passos2015}; since driver features generally denote optional features 
targeting end-user selection, the kernel itself is independent from them. In the case 
of \repo{torvalds/linux}, we exclude all driver-related code, which is, for the 
most part, inside the \tcode{driver} folder of the Linux kernel source code 
tree.\footnote{Specifically, we identify all driver-related code by executing a 
specialized script from G. Kroah-Hartman, one of the main developers of the 
Linux kernel. Available at 
\url{https://github.com/gregkh/kernel-history}.}

In addition to the Linux kernel repository, two other 
systems have a large amount of plugin-related code: \repo{Homebrew/home\-brew} 
and \repo{caskroom/homebrew-cask}. Homebrew is a package manager in Mac OS for 
handling the installation of different software systems. Its implementation 
allows contributors to push new formulas (automated installation recipes) to the 
system's remote repository, leading to thousands of formulas. As an extensible 
software system, Homebrew has one of the largest base of developers on GitHub 
(more than 5K developers, as of July 14$^\text{th}$, 2015). Considering all its 
formulas, Homebrew's TF, as computed by our heuristic, is 250. After excluding 
the files in folder \tcode{Library/Formula}, however, HomeBrew's truck factor reduces 
to 2. This clearly evidences the sensitivity of TF-values in the face of 
external code.  As for \repo{caskroom/homebrew-cask}, we ignore its
\tcode{Casks} directory.

In total, our list of ignorable files excludes 10,450 entries.\\[0.2cm]
\noindent\textit{Setting DOA Thresholds.} To find suitable 
thresholds, we manually inspect a random sample of 120 files stemming from the 
six top-most popular systems in our target corpus ($T^2$), one for each target 
language we account for. This results in files from \repo{mbostock/d3} 
(JavaScript), \repo{django/django} (Python), \repo{rails/rails} (Ruby), 
\repo{torvalds/linux} (C/C++), \repo{elasticsearch/ela\-stic\-search} 
(Java), and \repo{composer/composer} (PHP). We then compute the normalized 
DOA-values for each developer contributing at least one commit changing a file  
in our sample. Initially, we note that normalized DOA-values below 0.50 
lead to doubtful authorships. We measure doubtfulness by contrasting our 
authorship results with ranks we extract from \tcode{git-blame} reports. The 
latter contains the last developer who modified each line in a 
file~\cite{chacon2014pro}; 
by ranking developers according to the number of their modified 
lines, authors are likely to be those with higher ranks. Fixing 3.293 (which corresponds to the constant term in DOA's linear equation) as 
minimal absolute DOA and resetting the threshold for normalized DOA to 0.75 
better aligns results. 
Specifically, 64\% of the authors selected using those thresholds are 
classified as the top-1 ranked developer 
from \tcode{git-blame}; in 91\% of the cases, they 
are among the top-3 in the ranking list of \tcode{git-blame}, whereas 7\% 
lie between the 4$^\text{th}$ and 8$^\text{th}$ positions. In 
only three cases (2\%), the authors do not pair with any developer from 
\tcode{git-blame} rankings.

\subsection{Survey Design and Application} 

After collecting the truck factors of our chosen targets, one of the authors of 
this paper set to elaborate survey questions seeking to evidence the 
reliability of our results, as well as an instrument to get further insights. 
Following best practices in survey design~\cite{Shull2007}, we assure clarity, 
consistency, and suitability of our questions by running a feedback loop between 
the survey author and two other authors of this paper, until reaching a 
consensus among all three. We also perform a pilot study to identify early 
problems, such as whether our language correctly captures the intent of our 
questions. From the pilot study, we note few, but important communication 
issues, which we fix accordingly.\\[0.2cm]
\noindent\textit{Survey Questions.} After our pilot study, we phrase 
our questions as follows.\\[0.2cm]
\noindent\fbox{\begin{minipage}{.47\textwidth}
Question 1. Do developers agree that top-ranked authors 
are the main developers of their projects?
\end{minipage}
}\\[0.2cm]
This question seeks to assess the accuracy of our top authorship results. The 
top-ranked authors of a system are those we remove during the iteration step of 
our greedy-heuristic (recall Algorithm~\ref{alg:tf}), \emph{i.e.}, those responding for 
a system's truck factor. Note that we use the term \textit{main developers}, 
not authors. Our pilot study shows that developers tend to 
consider the creator of a file as its main author.\\[0.2cm]
\noindent\fbox{\begin{minipage}{.47\textwidth}
Question 2. Do developers agree that their project will be 
in trouble if they loose the developers responding for its truck factor?
\end{minipage}
}\\[0.2cm]
This question aims to validate our TF estimates. If we receive a 
positive feedback in this question, we can conclude that code authorship is an 
effective proxy.\\[0.2cm]
\noindent\fbox{\begin{minipage}{.47\textwidth}        
Question 3. What are the development practices and characteristics that can 
attenuate the loss of the developers responsible for a system's truck factor?
\end{minipage}
}\\[0.2cm]
Our intention here is to reveal the instruments developers see as most effective 
to circumvent the loss of  important developers---\emph{e.g.}, by devising better 
documentation, codification rules, modular design, etc.\\[0.2cm]
\noindent\textit{Survey Application.} Before contacting developers and applying 
our survey, we aim at calling their attention by promoting our work in popular 
programming forums (\emph{e.g.}, Hacker News) and 
publishing a preprint at 
PeerJ (\url{https://peerj.com/preprints/1233}).

After promotion, we apply the survey by 
opening GitHub issues in all target projects allowing such a feature (114 out of 133). The 
choice for issues is twofold: (i) issues foster public discussions among 
project developers; (ii) issues document all discussions, making them available 
for later reading. Potential readers include new developers, end-users 
interested on the target systems, researchers, etc. 

\begin{figure}[!t]
 \centering
 \includegraphics[width=.97\linewidth]{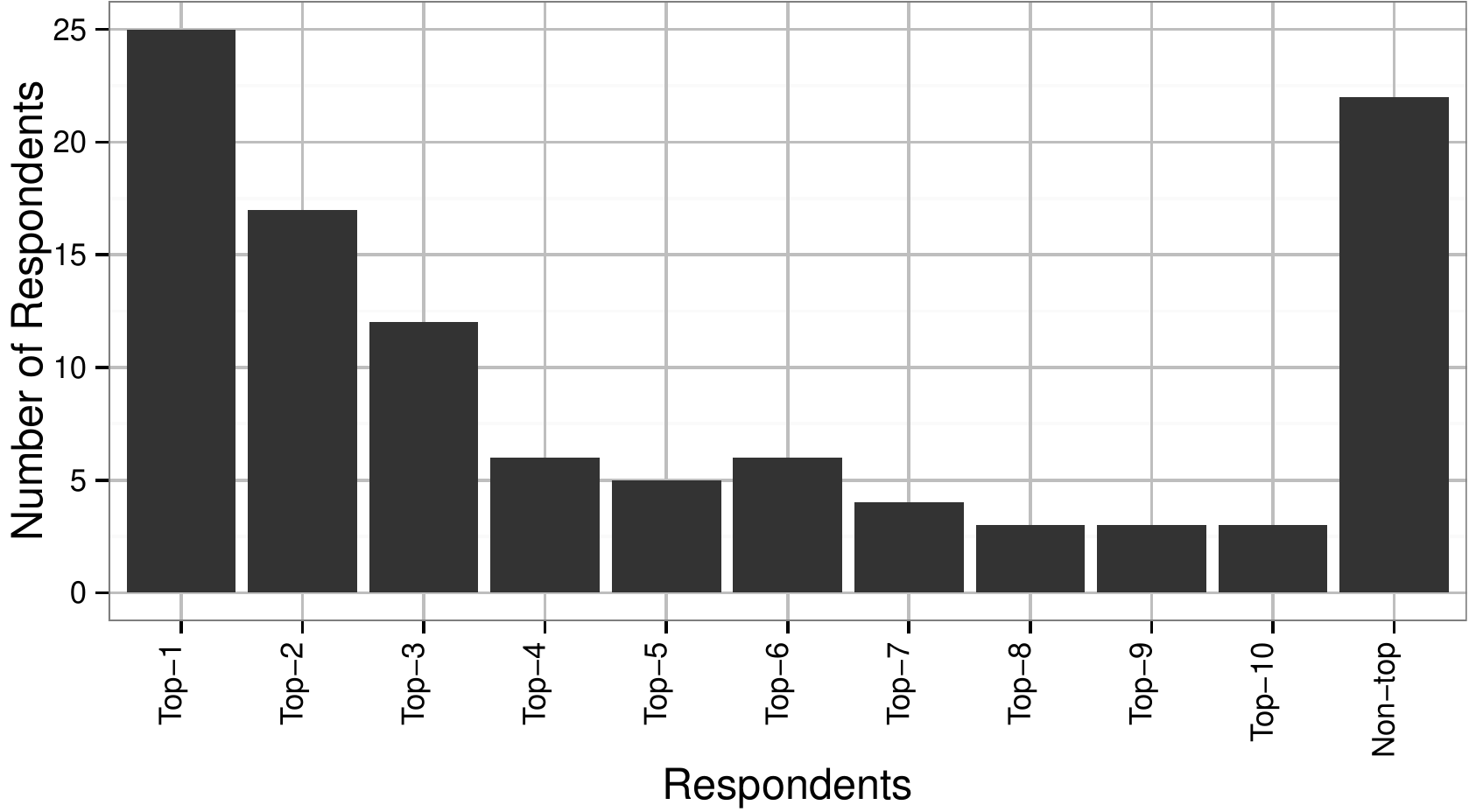}
 \caption{Respondents profile}
 \label{fig:respondents}
\end{figure}

Our posting period ranges from July 31th to August 11th, 2015. In the 
following two weeks, we set to collect answers, respected the deadline of August 
25th, 2015. In total, we collect answers from developers of 67 
systems. In 37 of those, there is a single answer from a single developer. 
However, often, the issues include discussions among different project 
members. For example, in \repo{saltstack/salt}, we have comments from six 
developers. In total, we accumulate 170 discussion messages from 106 
respondents; 96 messages stem (57\%) from the top-10 contributors of the 67 
participating projects. Figure~\ref{fig:respondents} characterizes all 
participants according to their level of project contribution. We get the list 
of top contributors by consulting the project's statistics as provided by 
GitHub. To exclude unreliable answers, we discard issues that do not have a single 
answer from a top-10 project contributor---five issues in total. Thus, we are 
left with 62 participating systems, as we have one issue per system. Among the 
issues that we do not exclude from analysis, we find 96 different respondents. 
Some messages are quite detailed. For instance, a message in an issue in 
\repo{elastic/elasticsearch} contains 1,670 words. In fact, according to the 
respondent, it triggered interesting internal discussions.\\[0.2cm]
\noindent\textit{Survey Analysis.} To compile the survey results, we analyze the 
discussions of our opened issues. For the first two questions, we classify 
answers according to four levels: agree, partially agree, disagree, or unclear.
The fourth level refers to cases where we cannot derive a clear position from  
an answer.  Two authors of this paper independently classified all answers, 
later crosschecking their results. 

As for our third question, we categorize answers to identify
common practices and characteristics. Our categorization closely
follows grounded theory open-coding principles~\cite{Straus1998}.

% \subsection{Performance Experiments}
% \todo[inline]{todo}

% . . . . . . . . . . . . . . . . . . . . . . . . . . .

\section{Truck Factor Estimates}
\label{sec:estimates}

\subsection{Preceding Output}

\begin{figure}[!t]
  \centering
  \includegraphics[width=0.8\linewidth]{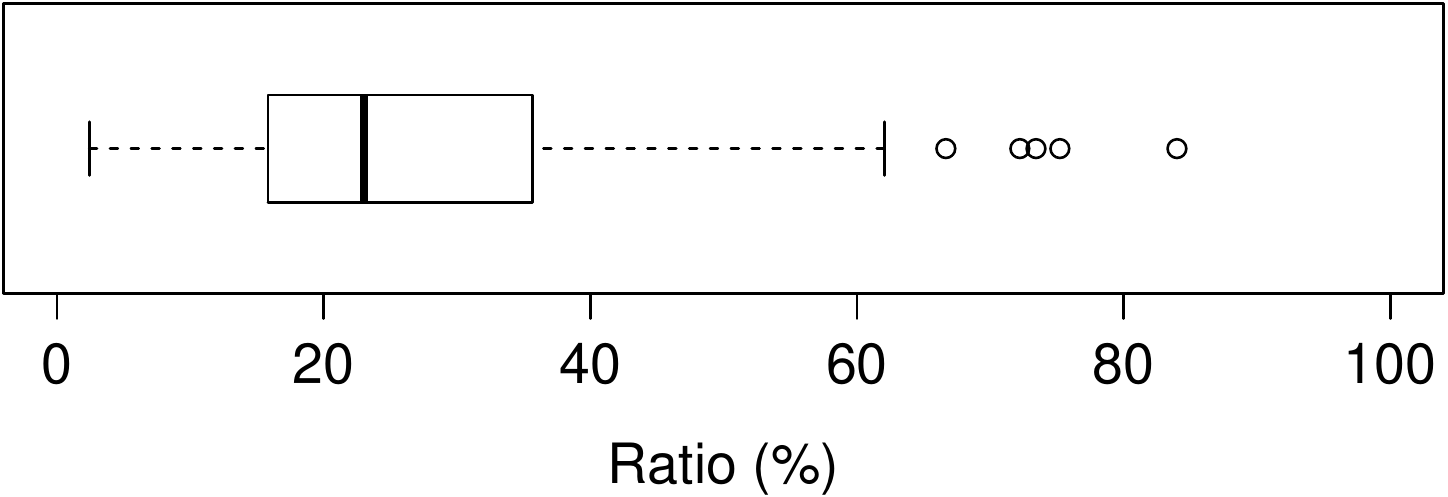}
  \caption{Proportion of developers ranked as authors}
  \label{fig:authorsratio}
\end{figure}

\noindent\textit{Target List of Source Files (Step 1).} Using our input list 
of ignorable files (see Section~\ref{sec:methodology}), as well as the 
automated exclusion by Linguist, we estimate the truck factor of 
243,660 files (33 MLOC)---$34\%$ less files than the original set in our 
subjects. The most frequent kind of files we remove 
concern JavaScript (5,125), PHP (3,099), and C/C++ (2,049) source 
files. Decreasing the number of target files decreases the target number of 
developers (63,193) and commits (1,262,130), a reduction of 28\% and 39\% 
w.r.t the original state of our target repositories.\\[0.2cm]
%
%
% \begin{figure}[tb]
%     \center      
%     \subfigure[Files]{\includegraphics[width=0.31\linewidth]
%         {images/boxplot-selected-files.pdf}
%         \label{fig:box-dataset:files}}
%     \subfigure[Commits]{\includegraphics[width=0.31\linewidth]
%         {images/boxplot-selected-commits.pdf}
%         \label{fig:box-dataset:commits}}
%     \subfigure[Developers]{\includegraphics[width=0.31\linewidth]
%         {images/boxplot-selected-devs.pdf}
%         \label{fig:box-dataset:devs}}
%     \caption{Target subjects (after Step 1)}
%     \label{fig:box-dataset}
% \end{figure}
%
\noindent\textit{Authorship List (Step 4).} By applying the normalized DOA 
to define the list of authors in each target system, as well as their authoring 
files, step 4 reveals the proportion of developers ranked as authors---see
Figure~\ref{fig:authorsratio}. For most systems, such proportion is relatively 
small; the first, second, and third quartiles are 16\%, 23\%, and 36\%, 
respectively. Interestingly, systems with a high proportion of authors usually 
have support of private organizations. Examples include four of the top-10 
systems with the highest author ratio among developers, such as \repo{v8/v8} 
(75\%), \repo{JetBrains/intellij-community} (73\%), \repo{WordPress/WordPress} 
(67\%), and \repo{Facebook/osquery} (62\%).  We also detect two language 
interpreters among the  top-10 systems: \repo{ruby/ruby} (72\%) and 
\repo{php/php-src} (59\%). At the other extreme, there are systems with a very 
low author ratio---\emph{e.g.}, \repo{sstephenson/sp\-rockets} (3\%) and 
\repo{jashkenas/backbone} (2\%). \repo{Backbone} is also an example, with only 
six authors amongst its 248 developers. These six authors 
monopolize 67\% of commits. A similar situation 
occurs with \repo{sprockets} (a Ruby library for compiling and serving web 
assets): although 61 developers associate to commits in the evolution history, 
95\% of commits come from two authors only; moreover, 27 developers respond 
for a single commit modifying a single line of code.

%Overall, as Figure~\ref{fig:workload} shows, a small number of individuals
%author most files in our corpus. In fact, we find a high concentration of files 
%in the hands of the top-10\% of authors, ranging from 45\% in 
%\repo{fzaninotto/Faker} to 100\% in \repo{getsentry/sentry}, 
%\repo{ratchetphp/Ratchet}, and \repo{ni\-co\-lasgramlich/AndEngine}. 
%Such data strengthens the problem of knowledge concentration in software 
%projects. 
%
%\begin{figure}[!ht]
%  \centering      
%  \includegraphics[width=\linewidth]{images/workload.pdf}
%  \caption{Percentage of files authored per author.  Each bar represents 
%a system and  the colors represent specific percentiles. Because a file can 
%have multiple authors, the bars of some systems exceed 100\%.}
%  \label{fig:workload}
%\end{figure}

\subsection{Results}

\begin{figure}[!t]
  \centering
  \includegraphics[width=0.8\linewidth]{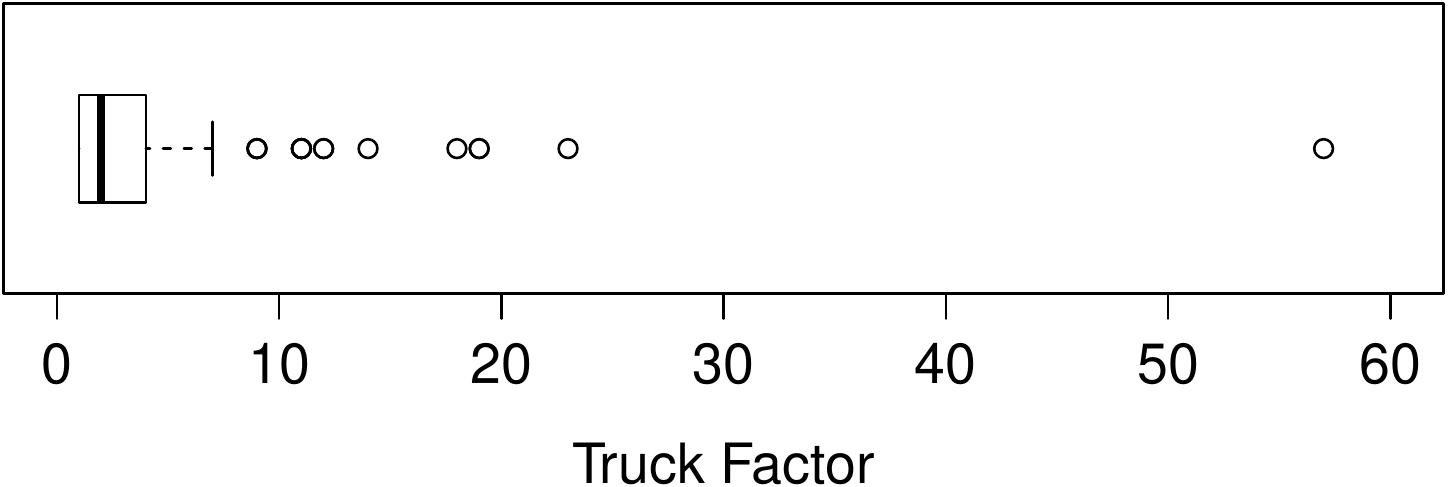}
  \caption{Systems Truck Factor}
  \label{fig:tf}
\end{figure}

Figure~\ref{fig:tf} presents the distribution of the truck factor amongst 
our subjects. The first, second, and third quartiles are 1, 2, and 4, 
respectively. 
Most systems have a small truck factor: 45 systems (34\%) have 
TF\,=\,1 (\emph{e.g.}, \repo{mbostock/d3} and \repo{less/less.js}); 
in 42 systems (31\%), TF\,=\,2 including well-known systems such as 
\repo{clojure/clojure}, \repo{cucumber/ cu\-cumber}, \repo{ashkenas/ backbone} 
and \repo{elasticsearch/e\-las\-ticsearch}.
Systems with high TF-values, however, do exist. Table~\ref{tab:tf} presents the 
top-15 (boxplot outliers)\footnote{In the boxplot, seven outliers have 
overlapping values, causing them to be plotted above an existing datapoint.}   
systems with the highest truck factors. Among those, \repo{torvalds/linux} has 
TF\,=\,57, followed by  \repo{fzani\-notto/Faker} 
(TF\,=\,23) and \repo{android/plat\-form\_frame\-works\_base} (TF\,=\,19). 
Other well-known systems include \repo{php/php-src} (TF\,=\,18), 
\repo{git/git} (TF\,=\,12), \repo{v8/v8} (TF\,=\,11), and \repo{rails/rails} 
(TF\,=\,9).

\begin{table}[!t]
    \centering
    \small
    \caption{Systems with highest truck factors}
    \label{tab:tf}
    \begin{tabular}{@{}lrl@{}}
        \hline
        \multicolumn{1}{c}{\bf System} & {\bf TF}   \\
        \hline
        \repo{torvalds/linux}   & 57 & \sbar{57}{57} \\
        \repo{fzaninotto/Faker} & 23 & \sbar{23}{57} \\
        \repo{android/platform\_frameworks\_base} & 19  &  \sbar{19}{57} \\
        \repo{moment/moment}    & 19 & \sbar{19}{57} \\
        \repo{php/php-src}      & 18 & \sbar{18}{57} \\
        \repo{odoo/odoo}        & 14 & \sbar{14}{57} \\
        \repo{fog/fog}          & 12 & \sbar{12}{57} \\
        \repo{git/git}          & 12 & \sbar{12}{57} \\
        \repo{webscalesql/webscalesql-5.6} & 11 & \sbar{11}{57} \\
        \repo{v8/v8}            & 11 & \sbar{11}{57} \\
        \repo{Seldaek/monolog}  & 11 & \sbar{11}{57} \\
        \repo{saltstack/salt}   & 11 & \sbar{11}{57} \\
        \repo{JetBrains/intellij-community} & 9 & \sbar{9}{57} \\
        \repo{rails/rails}      & 9  & \sbar{9}{57} \\
        \repo{puppetlabs/puppet}& 9  & \sbar{9}{57} \\        
        \hline
    \end{tabular}
\end{table}

%Our results suggests a correlation between Truck Factor and the number 
%of files in a system---see Figure~\ref{fig:tf-correlation}. The Spearman 
%correlation test reports such correlation to be moderate ($\rho=0.41$ 
%and $p$-$\mathit{value}< 1.012e^{-06}$). Counter-example outliers exist---e.g, 
%\repo{spring-projects/spring-framework}, the sixth largest system in our 
%corpus, has over 5,000 files, but low TF (2). In the opposite extreme lies 
%\repo{fzaninotto/Faker}, with over 400 files and a TF-value of 23, 
%the second highest we find.
%
%\begin{figure}[!ht]
%    \centering
%    \includegraphics[width=0.8\linewidth]{images/tf-correlation.pdf}
%    \caption{Systems Truck Factor}
%    \label{fig:tf-correlation}
%\end{figure}

\section{TF Validation: Surveying Developers}
\label{sec:validation:tf}

We present our survey results from our filtered set of issues and their 
underlying messages---we only account issues having at least one message from 
a top-10 project contributor. In total, the answers we analyze stem from 106 
respondents, of which 84 are top-10 contributors. The final number of 
participating systems is 62.
\\[0.2cm]
%
% %%%%%%%%%%%%%%%%%%%%%%%%%%%%%%%%%%%%%%%%%%%%%%
% SURVEY QUESTION 1
% %%%%%%%%%%%%%%%%%%%%%%%%%%%%%%%%%%%%%%%%%%%%%%
%
\noindent\textit{Question 1. Do developers agree that the top-ranked authors 
are the main developers of their projects?}\\[0.2cm]

\begin{table}[!h]
 \vspace*{-0.3cm}
 \caption{Answers for Survey Question 1}
 \centering \small
 \begin{tabular}{ c  c  c c}
 \hline
 {\bf Agree} & {\bf Partially} & {\bf Disagree} & {\bf Unclear}\\ 
 \hline
 31 (50\%) & 18 (29\%)  & 9 (15\%) & 4 (6\%) \\  \hline
 \end{tabular}
 \label{tab:rq1}
\end{table}
Table~\ref{tab:rq1} summarizes the answers for our first question. 
Respondents of 31 systems (50\%) fully {\em agree} with our
list of main developers. Example agreements:\\[-.2cm]

\noindent{\em``Yes, that's me.''}---developer from \repo{bjorn/tiled}.\\[-.2cm]

\noindent{\em``I think that it is a reasonable statement to make. They have 
contributed by far the most and paved the way for the rest of 
us.''}---developer from \repo{composer/composer}.\\[-.2cm] 

%\noindent{\em We agree and are aware of the risky bus factor for this project, 
%and are trying to improve it. (developer from {\sc 
%thoughtbot/paperclip})}\\[-.2cm]

%\noindent{\em Yes, this is true, but lately more developers started 
%contributing. (developer from {\sc SFTtech/openage})}\\[-.2cm] 

Developers of 18 systems (29\%) {\em partially agree} with our list of 
top-ranked authors. The main disagreement stems from the historical 
balance between older and newer developers:\\[-.2cm]

\noindent{\em``Yes and no, historically yes, currently no, a team has been 
picking up the activity, your analysis seems to be biased on capital (existing 
files) rather than activity (current commits).''}---developer from 
\repo{kivy/kivy}.\\[-.2cm] 

\noindent{\em``I would have added @DayS and @WonderCsabo as main 
developers.''}---developer from 
\repo{ex\-cilys/\-an\-droid\-an\-no\-ta\-tions}\\[-.2cm] 

\begin{figure}[!ht]
 \centering     
 \includegraphics[width=\linewidth]{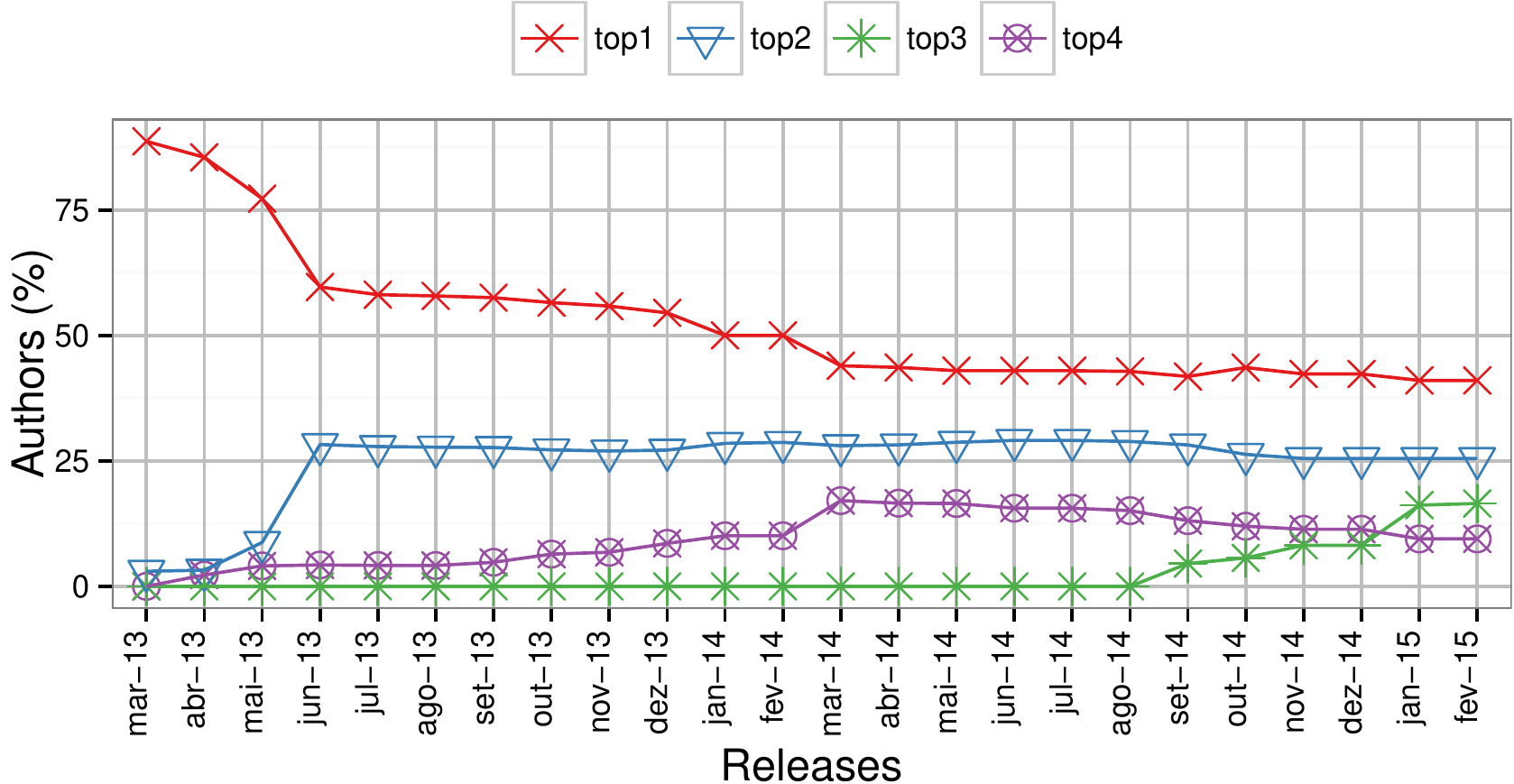}
  \caption{Percentage of files per author in 
  \repo{ex\-cilys/\-an\-droid\-an\-no\-ta\-tions} (top-4 authors)}
 \label{fig:percentage_files_top4}
\end{figure}

\noindent The latter answer illustrates a situation where we  
report two top-au\-thors in a target project; the respondent, although
agreeing with our suggestion, recommend adding two other developers.
The latter two have many recent commits; in contrast, one of the top 
authors we recommend is no longer active, strengthening the 
developer's argument. The two top-authors from our degree-of-authorship
measures cover 41\% and 26\% of files, respectively.
The two suggested by our respondent account for 9\% and 17\%  (see 
Figure~\ref{fig:percentage_files_top4}). However, we do note a gradual 
decrease in the number of authored files by the top developer we suggest, while 
an increasing trend for one of the two that our respondent recommends.
% Overall, we agree with 
% our respondents' inquiries.

Developers of nine systems (15\%) {\em disagree} with our list. Six developers 
indicate that other contributors are now responsible for their projects. 
Example disagreements include:\\[-.2cm]

\noindent{\em``No. TJ has been away from Jade for quite some time now. 
@ForbesLindesay is considered the main maintainer/develo\-per of 
Jade.''}---developer from \repo{jadejs/jade}.\\[-.2cm]

\noindent{\em``No, Burns hasn't been contributing for a while now. I've taken 
over what he was doing.''}---developer from \repo{backup/backup}.\\[-.2cm]

\noindent Other disagreements are due to auto-generated code. Finally, a single
developer has a negative attitude towards the question, providing us with no
insights.\\[0.2cm]
%
% %%%%%%%%%%%%%%%%%%%%%%%%%%%%%%%%%%%%%%%%%%%%%%
% SURVEY QUESTION 2
% %%%%%%%%%%%%%%%%%%%%%%%%%%%%%%%%%%%%%%%%%%%%%%
%
\noindent\textit{Question 2. Do  developers agree that their projects will be 
in trouble if they loose the truck factor authors?}\\[0.2cm]
Table~\ref{tab:rq2} summarizes results concerning this question. 

\begin{table}[ht]
  \vspace*{-0.3cm}
  \caption{Answers for Survey Question 2}
  \centering 
  \small
  \begin{tabular}{ c  c  c c}
  \hline
  {\bf Agree} & {\bf Partially} & {\bf Disagree} & {\bf Unclear}\\ 
  \hline
  24 (39\%) & 6 (10\%)  & 27 (43\%) & 5 (8\%) \\  \hline        
  \end{tabular}
  \label{tab:rq2}
\end{table}

Developers of 24 systems (39\%) {\em agree} with our truck factor results. 
Most positive answers are concise, usually a straight \aspas{yes} (14 answers). 
We consider as agreement answers that acknowledge a serious impact to the 
project if the given developers are to be absent, such as in:\\[-.2cm] 

\noindent{\em``If both of us left, the project would be kind of 
unmaintained.''}---developer from \repo{SFTtech/openage}.\\[-.2cm]

\noindent{\em``Initially, yes. However, given the size of the Grunt community, 
I believe a new maintainer could be found.''}---developer from 
\repo{gruntjs/grunt}. \\[-.2cm] 

\noindent{\em``If Wladimir or Pieter left, it would be a serious loss, but not 
fatal I think.''}---developer from \repo{bitcoin/bitcoin}.\\[-.2cm]

%Among the positive answers, we find a system that has in fact been
%discontinued---\repo{pockethub/Po\-cket\-Hub}. The project implements a GitHub 
%Android client, originally released as part of the GitHub platform. A GitHub 
%employee is identified as the system's single author, accounting for 78\% of 
%all source files. Recent commits are rare, dating back to 
%February 2014. One developer points out that there are only three people 
%working on the project, and they cannot fully dedicate to it as it is not 
%their full time job. 

Among the positive answers, we find a system that in fact \aspas{lost} 
its single truck factor author---\repo{pockethub/Po\-cket\-Hub}. The project 
implements a GitHub Android client, originally released as part of the GitHub platform. 
A GitHub employee is identified as the system's single author, accounting for 78\% of 
all source files. However, as stated in the repository home page, GitHub no 
longer maintains the app. The repository is almost inactive, 
receiving very few commits per month.
%, as indicated in Figure~\ref{fig:pockethub}. 
The last release dates from February 2014 (still as a GitHub project). 
One developer reports that low community involvement is the reason 
for the project's trouble, as there are only three people working on the 
project and this is not their full time job. 

% \begin{figure}[ht]
%   \centering
%   \includegraphics[width=\linewidth]{images/png/pockethub-graph.png}
%   \caption{Contributions to PocketHub}
%   \label{fig:pockethub}
% \end{figure}

Six answers are {\em partial agreements}. Examples:\\[-.2cm]

\noindent{\em``Somewhat agree. A loss in one area would mean a temporary dip 
in maintenance of that area until someone else stepped in.''}---developer from 
\repo{saltstack/salt}.\\[-.2cm]

\noindent{\em``Not necessarily, there's a long list of both small and 
significant contributors that had to understand a large piece of the code base 
to implement a feature or fix.''}---developer from 
\repo{justinfrench/formtastic}.\\[-.2cm]

Developers of 27 systems {\em disagree} with our TF-values. Six 
developers (22\%) have negative answers to our question, but do not 
provide further details; 21 developers (78\%) justify their answer stating 
that others could take over the project:\\[-.2cm]

\noindent{\em``Backup shouldn't be in trouble\ldots It's an open source 
project, anyone can start contributing if they want to.''}---developer from 
\repo{backup/backup}.\\[-.2cm]

We find two systems surviving the \aspas{lost} of the truck factor 
authors in our list:\\[-.2cm]

\noindent{\em``Coda was the author of the majority of the code. He left the 
project around a year ago. Some issues were going a long time without 
resolution, at which point I offered to maintain the project.''}---developer 
from \repo{dropwizard/metrics}.\\[-.2cm]

\noindent{\em``Your questions are timely, since Roland [the main author] has 
already left the project \ldots and we are not in trouble.''}---developer from 
\repo{caskroom/ho\-me\-brew-cask}.\\[-.2cm]

In the case of \repo{dropwizard/metrics}, it has been partially affected by 
the lost of its single truck factor author, as another developer was able to 
take over the project. As for \repo{caskroom/ho\-me\-brew-cask}, the respondent 
highlights two factors helping their transition after loosing their single 
truck factor author: (a) comprehensive documentation; (b) developers ready to 
transmit the rules and requirements to newcomers.\\[0.2cm]
%
% %%%%%%%%%%%%%%%%%%%%%%%%%%%%%%%%%%%%%%%%%%%%%%
% SURVEY QUESTION 3
% %%%%%%%%%%%%%%%%%%%%%%%%%%%%%%%%%%%%%%%%%%%%%%
%
\noindent\textit{Question 3. What are the development practices that can  
attenuate the loss of top-ranked authors?}\\[0.2cm]
Table~\ref{tab:questionc} summarizes answers. Documentation is the practice 
with the largest number of mentions across answers (36 answers), 
followed by the existence of an active community (15 answers), 
automatic tests (10 answers), and code legibility (10 answers). We 
group practices with a single answer under 
the \textit{miscellaneous} category---\emph{e.g.}, implementation in 
an specific programming language, periodic team chats, code reviews, support to 
classical algorithms, etc. In addition, we received seven \aspas{yes/no} 
answers, which are non-sensical given the nature of the question (not shown).

\begin{table}[!h]
  \centering
  \small
  \caption{Practices to attenuate the truck factor}
  \label{tab:questionc}
  \begin{tabular}{@{}lrl@{}}
  \hline
  \multicolumn{1}{c}{\bf Practice} & {\bf Answers}  & \\
  \hline
  Documentation                 & 36  & \sbar{36}{36} \\
  Active community              & 15  & \sbar{15}{36} \\
  Automatic tests               & 10  & \sbar{10}{36} \\
  Code legibility               & 10  & \sbar{10}{36} \\
  Code comments                 & 7 & \sbar{7}{36} \\
  Founding/Paid developers      & 5 & \sbar{5}{36} \\
  Popularity                    & 5 & \sbar{5}{36} \\
  Architecture and design       & 4 & \sbar{4}{36} \\
  Shared repository permissions & 4 & \sbar{4}{36} \\
  Other implementations         & 2 & \sbar{2}{36} \\
  Knowledge sharing practices   & 2 & \sbar{2}{36} \\
  Open source license           & 2 & \sbar{2}{36} \\
  Miscellaneous                 & 9 & \sbar{9}{36} \\
  \hline
  \end{tabular}
\end{table}

Although not development practices, having active communities and paid 
developers appear frequently among the answers we analyze. Both reasons appear 
as justifications for not concerning with top-authors lost:\\[0.2cm]
\noindent{\em``I'd say that the vibrant community is the reason for 
it.''}---developer from \repo{rails/rails}.\\[-.2cm]

\noindent{\em``We have a handful of other maintainers and a large body of 
contributors who are interested in Homebrew's future.''}---developer from 
\repo{Homebrew/homebrew}.\\[-.2cm]

%%%% Founding/Paid Developers %%%%%%%%%%%%%%

\noindent{\em``The people you listed are paid to work on the project, along 
with a number of others. So if the four of us took off, the project would hire 
some more people''}---developer from \repo{ipython/ipython}.\\[-.2cm]

%%% Architecture and design  %%%%%%%%%%%%%%

% % . . . . . . . . . . . . . . . . . . . . . . . . . . .

\section{Discussion}
\label{sec:discussions}

In this section, we discuss the lessons we learn in our study. We also lay 
out directions for future research on truck factor measurements and 
applications.

\subsection{DOA Results} 

The results produced by the DOA model seem accurate when applied 
to a large collection of systems. For the first survey question, the 
developers of 49 systems (84\% of the valid answers) agree or partially agree 
with our results. Despite that, some developers report that the model 
gives high emphasis on first authorship (FA) events. 
In the same question of our survey, six developers disagree with 
our results exactly due to this resilience of the DOA model in transferring
authorship from the first author to another one.  This applies in 
systems where a single developer creates the bulk of the code, but later 
switches role (\emph{e.g.}, project leader or mentor), becoming less active in 
development activities. In fact, some developers suggest that  DOA 
computation should consider only the most recent development history, \emph{e.g.}, 
commits performed in the last year. One developer from 
\repo{clojure/clojure} explicitly declares that \aspas{\textit{if the code is 
old enough, even the original author will have to approach it with essentially 
fresh eyes.}}

%\subsection{Read Events} 
%
%The DOA model only considers changes to infer authorship on a file. However, at 
%least expertise on the file can be acquired by just reading the code, without 
%touching it.
%One developer from {\sc emberjs/ember.js} mentioned that the methodology 
%followed in the study \aspas{overweights edits, which is understandable since 
%they are measurable.} A developer from {\sc clojure/clojure} had a similar 
%opinion: \aspas{I think the correct measure for maintainability is not degree 
%of 
%authorship, but code understanding.  Unfortunately, this is harder to measure.} 
%Interestingly, we did not receive a single answer in the survey reporting a 
%missing author that understands a lot about the code, but does not change it 
%frequently.
%Most answers complaining about missing developers usually reference their 
%recent 
%commits to the system, sometimes explicitly showing the commit activity graphs. 
%Anyway, to consider data about reads, the model should be augmented with a 
%measure like the degree-of-interest (DOI)~\cite{Fritz2010, Fritz2014}. However, 
%this would require a customized IDE to capture this information (like the Mylyn 
%Eclipse plug-in~\cite{KerstenM06,Murphy2006}). 

\subsection{Challenges on Computing Truck Factors}

We receive answers for 67 (out of 114) systems. This high response ratio (59\%) 
is certainly a consequence of the importance that developers give to the truck 
factor concept. By analyzing the answers, we see that developers generally 
recognize the impact that the truck factor may have in the public reputation of 
their systems. However, it is worth noting that estimating this concept 
automatically has many challenges. A few developers refused to answer our 
question, stating for example that \aspas{\textit{it is an existential, 
speculative question that I will not attempt to answer}} (developer from 
\repo{mbostock/d3}).
A second developer states  that \aspas{\textit{the truck factor is mostly 
concerned with institutional memory getting lost. No automatic system can 
account for this lost, unless all project communication is public.}} (developer 
from \repo{libgdx/libgdx}). Our survey also reveals that developers usually 
consider documentation as the best practice to overcome a truck factor 
episode.

Despite the challenges in computing truck factors automatically, our 
code-authorship-coverage heuristic presents compelling results. We 
receive positive or partially positive answers for 30 systems (53\% of the 
valid answers). Even when developers do not agree with our estimation, it 
is not completely safe to discard a possible damage to the system. For example, 
six developers state that truck factor is not a concern in open source 
systems, since it is always possible to recruit new core developers from their 
large base of contributors. In fact, we observe a successful transition of 
core developers in at least two systems. In contrast, we cannot discard the 
risks inherent to such transitions, specially when they should take place due to 
a sudden and unplanned truck-factor-like episode.

Developers also point two concrete problems in our heuristic for
computing truck factors. First, it considers all files in a system as equally 
important in terms of the features they implement. However, not all 
requirements and features are equally critical to a system survivability. We 
address this problem by discarding some files from our analysis, in the cases 
they lead to highly skewed results (\emph{e.g.}, recipes from 
\repo{Homebrew/home\-brew}). However, in other systems this partition between 
core and non-core files is less clear (at least, to non-experts). Second, the 
heuristic does not account the last time a file is changed. In the 
survey, some developers claim that losing the author of a very stable file is 
not a concern (since they probably will not depend again on this author to 
maintain the file). When such files are common in a system, the heuristic 
can be adapted to just consider recently changed files.

% % . . . . . . . . . . . . . . . . . . . . . . . . . . .

\section{Threats to Validity}
\label{sec:threats}

\noindent\textit{Construct Validity.} We compute the degree-of-authorship 
using wei\-ghts derived for other systems~\cite{Fritz2010, Fritz2014}. 
Therefore, we cannot guarantee these weights as the 
most accurate ones for assessing authorship on GitHub projects. However, the 
authors of the DOA formula show that the proposed weights are robust enough 
to be used with other systems, without computing a new regression. Still, we 
mitigate this threat by initially inspecting the DOA results for 162 pairs of 
authors and files. Contrasting results with those from \tcode{git-blame} 
suggest DOA-values to be reliable. 

The presence of non-source code files, third party libraries, and developers 
aliases can also impact our results. To address these threats our tool 
performs file cleaning and alias handling steps before calculating truck factor 
estimates.\\[0.2cm]
\noindent\textit{Internal Validity.} Our approach computes the authors of a 
file by considering all the changes performed in the target file. 
Therefore, our approach is sensitive to loss of part of the development history 
as result of a erroneous migration to GitHub. We mitigate this threat using a 
heuristic to detect systems with clear evidence that most of its development 
history was performed using another version control platform and that this 
history could not be correctly migrated to GitHub. Moreover, the full 
development history of a file can be lost in case of renaming operations, copy or file split (\emph{e.g.}, as result of a refactoring operation like \emph{extract 
class}~\cite{Fowler1999}). We address the former problem using Git facilities 
(\emph{e.g.}, \tcode{git log --find-renames}). However, we acknowledge the need for 
further empirical investigation to assess the true impact of the other 
cases.\\[0.2cm]
\noindent\textit{External Validity.}
We carefully select a large number of real-world systems 
coming from six programming languages to validate our approach.
Despite these observations, our findings---as usual in empirical software engineering---cannot be directly generalized to other systems, mainly closed-source ones.
Many others aspects of the development environment, like contribution policies, automatic 
refactoring, and development process may impact the truck factor results 
and it is not the goal of this study to address all of them.

Finally, to assess the impact of the aforementioned threats in our results, we 
conduct a survey with developers of the systems under analysis in this paper, 
as reported in Section~\ref{sec:validation:tf}. 

% . . . . . . . . . . . . . . . . . . . . . . . . . . .

\section{Related Work}
\label{sec:relatedwork}

Although widely discussed among {\em eXtreme Programming} (XP) practitioners, 
there are few studies providing and validating truck factor measures for a 
large number of systems.
Zazworka et al.~\cite {Zazworka2010} are probably the first to propose a formal 
definition for TF, specifically to assess a proj\-ect's conformance to XP 
practices. 
For the purpose of simplicity, their definition assumes that all developers who 
edi\-t a file have knowledge about it.
Furthermore, they only compute the TF for five small projects written by 
students.  Ricca et al.~\cite{Ricca2010, Ricca2011} use Zazworka's 
definition to compute truck factors for opensource projects. In their first 
work, they propose the use of the TF algorithm as strategy to  identify 
\aspas{heroes} in software development environments. 
In their second work, the authors point for scalability limitations in
Zazworka's algorithm, which only scales to small projects ($\leq$ 30 
developers). In our study, 122 out of 133 systems have more than 30 developers 
(maximum is \repo{torvalds/linux}, with thousands of developers among non-driver 
files).
%for problems of the TF measurement, in which the number of modifications is not 
%considered and all developers who performs a file commit have the same 
%importance, and the need of further validation of TF measurements.
Hannebauer and Gruhn~\cite{Hannebauer2014} further explore the scalability 
problems of Zazworka's definition, showing that its implementation  
is NP-hard. Cosentino et al.~\cite{Cosentino2015} propose a tool to calculate TF 
for Git-based repositories. They use a hierarchical strategy, aggregating 
file-level authorship results to modules and, in a second step, 
aggregating module-level results into systems. They evaluate their tool with 
four systems developed by members of their research group.

Overall, our study differs from the previous ones in three main points: we 
use the DOA model to identify the main authors of a file; we evaluate our 
approach in a large dataset composed of real-world software from six programming 
languages; we validate our results with expert developers.

%GitHub studies

% . . . . . . . . . 	. . . . . . . . . . . . . . . . . .

\section{Conclusion}
\label{sec:conclusion}

This paper proposes and evaluates a heuristic-based approach to estimate a 
system's truck factor, a concept to assess knowledge concentration among team 
members. We show that 87 systems (65\%) have $\mathit{TF} \leq 2$. 
We validate our results with the developers of 67 systems. 
In 84\% of the valid answers, respondents agree or partially agree that the 
TF's 
authors are the main authors of their systems; in 53\% we receive a positive or partially 
positive answer regarding the estimated truck factors. 
 
According to the surveyed developers, documentation is the 
most effective development practice to overcome a truck factor event, followed 
by the existence of an active community and automatic tests. We also comment 
on the main lessons we learn from the developers' answers to our questions.

As future work, we plan to introduce and evaluate the improvements suggested by the 
surveyed developers in our heuristic to compute truck factors. As an example, 
we can consider only recently changed files in the estimation of TFs and compute
authorship at line level.
Finally, we intend to perform a second study, considering industrial and closed 
systems, and compare the truck factor results  with the ones we report in this 
paper for opensource systems.

\section*{Acknowledgments}

\noindent We thank all the respondents of our survey.
This study is supported by grants from FAPEMIG, CNPq, and UFPI.

% . . . . . . . . . . . . . . . . . . . . . . . . . . .

\bibliographystyle{IEEEtran}
\bibliography{references}

% Fix indentation problem in the last line of the paper
\par\leavevmode

\end{document}